\documentstyle[11pt,newpasp,twoside,epsf]{article}
\renewcommand{\vec}[1]{\mbox{\boldmath $#1$}}
\begin{document}
\title{Pulsar Electrodynamics}
\author{Shinpei Shibata}
\affil{Department of Physics, Yamagata University, JAPAN}
\begin{abstract}
First I emphasize the magnetospheric
loop current system induced by the wind activity.
Next, I address the formation of the field-aligned electric field.
An interesting aspect is the
reaction of the inner magnetosphere to an induced current.
In the traditional outer gap, the current running through the gap
is all carried by the particles created in the gap. However,
if external current-carrying particles are added, the gap moves
inward or outward;  it can even appear near the star.
It is suggested that the gaps appear at various altitudes, depending on
the current distribution on field lines, the field line geometry,
and the source of the current-carrying particles.

For definiteness of sign of charge and current direction,
$\vec{\Omega} \cdot \vec{\mu} >0$ is assumed throughout.
\end{abstract}

\section{Magnetospheric Current System}

\subsection{Pulsar Wind}
The rotational energy of pulsars is carried off mostly by the pulsar wind.
Pulsed radiation only accounts for a small fraction of the rotation power.
At least for young pulsars,
this idea is supported by observations of pulsar powered nebulae.
{\it The pulsar wind is predominant in the pulsar electrodynamics.}

However, the nature of the pulsar wind is not clear.
The wind is conventionally
thought to be an outflow of magnetized plasmas, which transports
energy in the forms of a Poynting flux and a kinetic energy flux.

Dominance of the wind may not be the case
if the electromotive force of the star
is marginally reduced to the voltage required for pair creation.
Nevertheless,
we assume dominance of the wind in the following,
and this is hopefully justified for most pulsars.

\subsection{Electric current in the magnetosphere}
For the axisymmetric case,
if there is an outflow of energy, then a simple result
is a loop current system.
As shown in Figure~1 (left),
the current starts from the star,
goes out and returns back to the star on different field lines.
A DC circuit is formed, connecting the central dynamo and the load (the wind
region beyond the light cylinder).
The poloidal electric field produced by the dynamo and the toroidal
magnetic field
by the loop current
make the outward Poynting flux.

\begin{figure}
\begin{center}
\mbox{ \epsfxsize=5cm \epsfbox{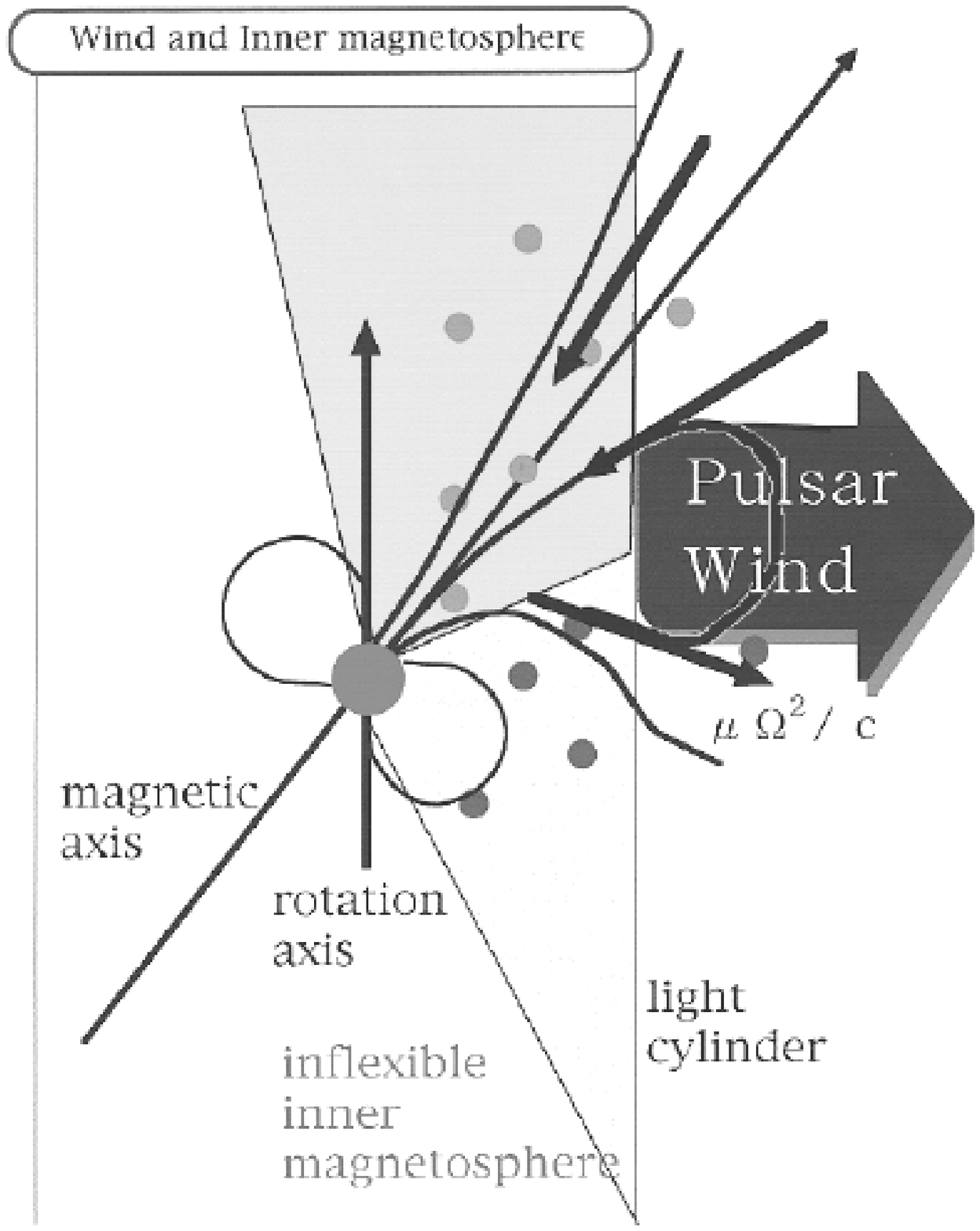}
\epsfxsize=5cm \epsfbox{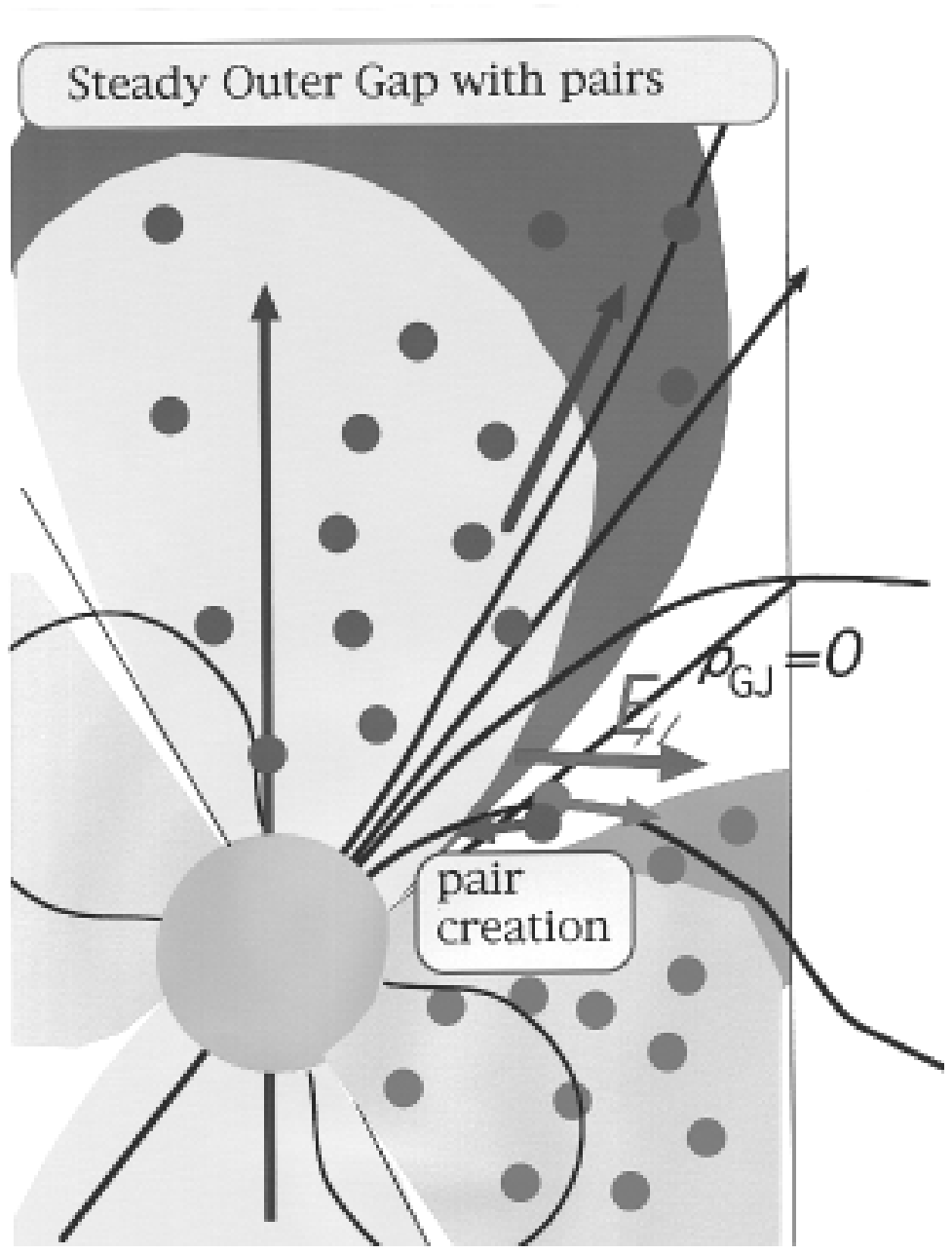} }
\end{center}
\caption{Left: The energy release from the magnetosphere is
associated with the loop current system.
Right: If pairs are created in the gap, then corotation regions expand
to fill the space within the light cylinder. The associated current system
is compatible with the wind activity on the left.}
\end{figure}

The current can go away to infinity and return back from infinity.
However, if some part of the current closes somewhere
in the outer magnetosphere, then
Poynting flux is converted into the kinetic energy of the plasma
to accelerate the wind ($\vec{E} \cdot \vec{j} >0$).
How much current closes in the wind is the issue
whether kinetic dominant winds are formed or not.
To my knowledge, this is still controversial.

For oblique cases, a non-zero displacement current
complicates the interpretation.
However, it can be shown that
if there is {\em no field-aligned electric field}
($E_\parallel \equiv \vec{E} \cdot \vec{B} / B =0$),
the outward energy flux {\em does require a `real' loop current}
even with displacement current.
Only for a non-vanishing field-aligned electric field
($E_\parallel \neq 0$) does the displacement current contribute
to the outflow of energy, as in the case of
magnetic dipole radiation
(cf.\ the Appendix of Shibata
\& Hirotani, 1999).

An illustrative example is the oblique force-free model
without field-aligned current. In this case, there is no
Poynting flux across the light cylinder
(Henriksen, \& Norton 1975;
Beskin, Gurevich, \& Istomin 1993;
Mestel, Panagi, \& Shibata 1999)

If the potential drop associated with probable $E_\parallel \neq 0$
is much smaller than the available voltage,
then in general, the
`real current loop' dominates in determining
the rotational energy loss.
As a result,
the field-aligned current with an intensity of order of the Goldreich-Julian
(GJ) value,
is forced to run through the inner magnetosphere,
connecting the generator, the neutron star, and the main load, which is the
wind.
This view seems to be applicable at least for young pulsars because
their electromotive force is large enough to make electron-positron pairs.

\section{Constraints for the inner magnetosphere}

The inner magnetosphere is {\it inflexible}:
\\
(1) the magnetic field is hardly changed by the GJ
current.  It is essentially current free, say, dipolar.
\\
(2) the order of strengths of various forces, i.e.,
pressure force $\ll$ gravitational force $\ll $
electromagnetic force,
results in
(i) it is the electromagnetic force that controls the particle motion;
(ii) due to strong gravity, a quasi-neutral plasma cannot be supported
above scale heights (a few cm).
Quasi-static plasmas should be completely
charge separated:
plasmas are composed of either electrons only or ions/positions only;
(iii) pair plasma can stay quasi-neutral only if the field-aligned electric
field in it is screened out.
\\
(3) finally the smallness of particle inertia requires the ideal-MHD
condition,
$\vec{E} + \vec{v} \times \vec{B}/c =0$.
If this applies,
the particles follow the corotation plus
field-aligned motion:
$\vec{v} = \vec{u}_{\rm c} + \kappa \vec{B}$,
where $\vec{u}_{\rm c}=\vec{\Omega} \times \vec{r}$ is the corotation velocity,
and the electric field becomes the corotational electric field,
$\vec{E} = - (1 / c) \vec{u}_{\rm c} \times \vec{B} \equiv \vec{E}_{\rm c}$,
which is supported by the GJ charge density
$\rho_{\rm GJ} =  \nabla \cdot \vec{E}_{\rm c}/  4 \pi$.
Near the star, the general relativistic effect changes the form of
$\rho_{\rm GJ}$, which in turn
changes the acceleration field for a given current density.
The corotational electric field has no component along the magnetic
field, $E_\parallel =0$.

Although the
particles nearly follow the rigid field lines,
ideal corotation  is not realized
everywhere in the
inner magnetosphere.
This is because the
particle supply does not lead to
the GJ density everywhere.
Even a small deviation of the real charge density from the
GJ charge density creates a field-aligned electric field.
The unscreened electric force is balanced by an inertial force;
oscillations (Langmuir type) can develop and,
in some cases, a huge potential drop appears, which can be large enough
for renewed electron-positron pair creation,
depending on situations, e.g.,
imposed current density, field geometry, emission from
the neutron star.

The nature of completely charge separated flows
depends on field line curvature.
The charge density of the flow
varies according to the equation of continuity,
$\rho_e = e n \propto B/v$,
and thereby once the flow is relativistic,
the charge density varies in proportion to the magnetic field strength:
$\rho_e \propto B$.
On the other hand, the GJ density changes as
$\rho_{\rm GJ} \propto B_{\rm z}$.
Hence, even if the GJ density is realized somewhere
in the flow, the charge density deviates from the GJ value and
$E_\parallel$ should appear.
On field lines curving toward the rotation axis,
the ratio, $\rho_e / \rho_{\rm GJ} = B/B_{\rm z}$,
decreases outwardly
so that the charge tends to be depleted on these field lines.
On field lines curving away from the rotation axis,
the charge tends to exceed the GJ value.
In some regions, when the flow is non-relativistic,
the GJ density appears  for a long distance
by adjusting the velocity so as to screen the field-aligned
electric field.
Such a flow in general has spatial oscillation of the Langmuir type.

\section{Outer Gap}
\subsection{Quasi-Static Consideration}

The original idea of the outer gap is as follows (Holloway  1973).
Suppose some particles are ripped off through the light cylinder
due to the centrifugal force, wave pressure or whatever.
Particles in the inner magnetosphere become insufficient
to provide the GJ charge density. The density perturbation grows and
the charge separation by the central dynamo is so strong
that the region around the {\em null surface} where $\rho_{\rm GJ} =0$ is left
vacant with unscreened $E_\parallel$.

Refilling particles back through the l-c is unlikely because
it opposes the emf.
More likely, additional charges are provided by an
electron-positron pair creation cascade in the gap,
perhaps initiated by a cosmic ray in the gap.
The created pairs are immediately charge-separated so as to
refill the charge depleted regions.
The gap will then be reduced.
Furthermore,
electrons going back to the star give negative charge to the star, and
as a result
the vacant space above the polar dome can even be refilled.

After all, the corotation region expands and occupies the space
within the light cylinder as far as possible.
The expansion of the corotation region may induce further loss of particles.
More importantly, the refilling action produces an electric current system.
It is outward across the outer gap and inward in higher latitudes, see
Figure~1 (right).
This current is completely consistent with that required by the wind.
Thus it is indicated that
the pair creation to refill the gap cooperates with the wind:
the outer gap accelerator will operate in the loop current steadily.
A bonus is for the wind to get the quasi-neutral plasmas via
pair creation.

\subsection{A Steady Outer Gap Model with Pair Creation}

Hirotani \& Shibata (1999a, 1999b, 1999c) calculated the electric field
in a steady one-dimensional outer gap
self-consistently with a gamma-ray distribution
function and flows of electrons and positrons created
by photon-photon collision.
The gap is partially refilled, but still has a field-aligned
electric field.

The gap model is somewhat similar to a semiconductor in the sense
that the concept of holes and real charged particles is convenient.
There are effectively positive holes to the left of the null surface, and
negative holes to the right of the null surface in Fig.\ 1.
The charge density of the holes is minus the GJ
value, $- \rho_{\rm GJ}$.
For example,
on the left, if the positive holes are filled with electrons, then
the real negative charge equals the GJ value and
the field-aligned electric field can be screened out
$\rho_e - \rho_{\rm GJ} = 0$.
In the gap, two regions with oppositely charged holes are facing each
other, and a field-aligned electric field appears.

Particles are accelerated in the gap and emit gamma-rays, which
collide with soft photons to make pairs.
The pairs are immediately separated in opposite directions.
These pairs produce space charge in the gap so as to refill the gap in part.
The field aligned electric field is weakened.

These processes are described by
the Poisson equation, equations of motion, equations for
the gamma-ray distribution function with source and sink terms,
and equations of continuity for electrons and positrons again with a source
term.
HS solved these equations for a steady outer gap.

\subsection{A Steady Outer Gap Model with external current sources}

Conventional outer gap models assume that all the current
carrying particles are created in the gap.
However, external current-carrying particles  may come into the gap:
current-carrying
electrons may come in from the wind, or positrons may be emitted
from the stellar surface.

This external current supplies an additional space charge
as if a back ground charge is added to the GJ charge density.
As a result,
the position of the outer gap moves inward or outward.
In principle, the external-current-dominated outer gap can
appear near the star.
In this sense there is no obvious distinction between the outer gap and
polar gap. The difference is just one parameter that is
the ratio of the {\em external} current-carrying particles to the
current-carrying
particles {\em produced in the gap}.
In contrast to the outer gap, conventional polar cap models assume a completely
external current source.

\section{Models of the inner magnetosphere}

Let us summarize the various reactions of the inner magnetosphere
to the imposed current.
The inner magnetosphere through which electric current is
forced to run can be classified according to the field line curvature
and the direction of the current.
The direction and intensity of the current on each field line
are determined globally,
and are more like quantities imposed by the wind at least for
young pulsars.


The outer gap is located on field lines with away curvature and an
outgoing current.
As has been described in detail, a steady, pair-creating outer
gap is possible.
If the current running through the gap is all carried by
electrons and positrons {\em created in the gap}, the
gap is located at the null surface.
If some part of current is carried by externally supplied
particles, then
the gap shifts its position, but
the electrodynamics itself is the same as that which applies
around the null surface.

As for the polar cap accelerators,
on field lines curving `toward' the rotation axis, the
Arons type model is known
(Scharlemann, Arons, \& Fawley 1978;
Arons, \& Scharlemann 1979;
Shibata 1997),
where an external super GJ current
is assumed to be a critical value.
The region near the star is over-filled with negative charge,
while the down stream region becomes charge-depleted due to
field line curvature ($B/B_{\rm z}$ decreases).
Therefore for these two regions, one needs positive charge and the other
needs negative charge in order for the real charge density
to adjust to the respective GJ values.
Thus the electrodynamics of this accelerator is as same as the outer gap.
The difference is that the polar gap current is externally supplied.
The gap can make pairs, and some of the pairs contribute to the
current so that the polar gap model can be modified
with larger current densities than the critical value.
In this case, one has a flux of backward positrons.

However, if the current density required by the wind is smaller
than a certain value, the flow is an oscillatory GJ flow without
any acceleration (Shibata 1997).

Acceleration takes place  also on field lines curved away
from the rotation axis. In this case, again the
electrons are assumed to flow out.
Even if the flow is initially an oscillatory GJ flow near the star,
as the flow goes up,
{\em the space automatically becomes over-filled} at some distance
because of the field line curvature ($B/B_{\rm z}$ increases).
This catastrophic break up of an $\vec{E} \cdot \vec{B} =0$ condition
may be terminated by forming
a positive charge region downstream.
This is possible by polarization of a pair plasma.
Electrons are accelerated to emit gamma-rays and subsequently
to make pairs. Pair positrons are decelerated, and as a result
positive space charge appears.

However, pair creation rate for this screening to operate
is much higher
than the value expected by the normal magnetic pair creation:
the simple idea that once the pair density exceeds
the GJ density, the field-aligned electric field is immediately
screened out is not correct
(Shibata, Miyazaki, \& Takahara 1998).
The reaction of the away curved field lines to the imposed current
has yet to be clarified.

\section{Future}

As seen above, the inner magnetosphere reacts to the imposed current
in various ways
depending on current direction, current  density,
field geometry, external particle flux, and so on.
The accelerators can appear at various altitudes,
so there is no clear discrimination between the outer gap and the polar gap.
This might be confirmed observationally by the identification
of various new pulse components.
We need more accurate local models of the field-aligned accelerators,
which should predict the properties of the high energy emission.
The local models should have  adjustable free parameters, such as
current density and external particle flux, and should be self-consistent
in the sense that the electric field,  the pair production process, the
photon distribution and motion of the particles
are all solved together.
Such sophisticated models can only interpret detailed
observations giving phase (component) resolved energy spectra.


\begin{references}
\reference Arons, J., Scharlemann, E.T., 1979, \apj 231, 854
\reference Beskin V. S., Gurevich A. V., \& Istomin Ya. N., 1993, ``Physics
of the
Pulsar Magnetosphere'', Cambridge University Press, page 163
\reference Henriksen R. N., \& Norton J. A., 1975, \apj, 201, 719
\reference Hirotani K., Shibata S., 1999a, \mnras 308, 54
\reference Hirotani K., Shibata S., 1999b, \mnras 308, 67
\reference Hirotani K., Shibata S., 1999c, \pasj 51, 683
\reference Holloway,  N.J., 1973, Nat Phys. Sci., 246, 6
\reference Mestel L., Panagi P., Shibata S., 1999, \mnras 309, 388
\reference Scharlemann, E.T., Arons, J., Fawley, W.M.,  1978, \apj 222, 297
\reference Shibata S., \& Hirotani K., 1999, in preparation
\reference Shibata, S.,  1997, \mnras 287, 262
\reference Shibata, S., Miyazaki, J., Takahara, F., 1998, \mnras 295, L53
\end{references}
\end{document}